\documentclass[pra,aps,amsmath,amssymb,amsfonts,onecolumn,nofootinbib]{revtex4}
\usepackage{amssymb}
\usepackage{}
\usepackage{bm,mathrsfs}

\usepackage{graphicx}
\usepackage{epsfig}
\usepackage{amsmath,bbm}
\usepackage{amsfonts,amssymb}
\usepackage{times}
\usepackage{verbatim}
\usepackage[sort&compress]{natbib}

\usepackage{amsmath}
\usepackage[colorlinks,breaklinks,linkcolor=blue,anchorcolor=blue,citecolor=blue,urlcolor=blue,
dvipdfm
]{hyperref}

\newcommand{\be}{\begin{equation}}
\newcommand{\ee}{\end{equation}}
\newcommand{\bea}{\begin{eqnarray}}
\newcommand{\eea}{\end{eqnarray}}

\def\>{\rangle}
\def\<{\langle}

\def\qed{\leavevmode\unskip\penalty9999 \hbox{}\nobreak\hfill
     \quad\hbox{\leavevmode  \hbox to.77778em{%
               \hfil\vrule   \vbox to.675em%
               {\hrule width.6em\vfil\hrule}\vrule\hfil}}
     \par\vskip3pt}
     
\begin{document}

\newtheorem{theorem}{Theorem}
\newtheorem{lemma}[theorem]{Lemma}
\newtheorem{corollary}[theorem]{Corollary}
\newtheorem{proposition}[theorem]{Proposition}
\newtheorem{definition}[theorem]{Definition}
\newtheorem{example}[theorem]{Example}
\newtheorem{conjecture}[theorem]{Conjecture}
\title{Engineering steady Knill-Laflamme-Milburn state of Rydberg atoms by dissipation}

\author{Dong-Xiao Li}
\affiliation{Center for Quantum Sciences and School of Physics, Northeast Normal University, Changchun, 130024, People's Republic of China}
\affiliation{Center for Advanced Optoelectronic Functional Materials Research, and Key Laboratory for UV Light-Emitting Materials and Technology
of Ministry of Education, Northeast Normal University, Changchun 130024, China}

\author{Xiao-Qiang Shao\footnote{shaoxq644@nenu.edu.cn}}
\affiliation{Center for Quantum Sciences and School of Physics, Northeast Normal University, Changchun, 130024, People's Republic of China}
\affiliation{Center for Advanced Optoelectronic Functional Materials Research, and Key Laboratory for UV Light-Emitting Materials and Technology
of Ministry of Education, Northeast Normal University, Changchun 130024, China}

\author{Jin-Hui Wu}
\affiliation{Center for Quantum Sciences and School of Physics, Northeast Normal University, Changchun, 130024, People's Republic of China}
\affiliation{Center for Advanced Optoelectronic Functional Materials Research, and Key Laboratory for UV Light-Emitting Materials and Technology
of Ministry of Education, Northeast Normal University, Changchun 130024, China}

\author{X. X. Yi}
\affiliation{Center for Quantum Sciences and School of Physics, Northeast Normal University, Changchun, 130024, People's Republic of China}
\affiliation{Center for Advanced Optoelectronic Functional Materials Research, and Key Laboratory for UV Light-Emitting Materials and Technology
of Ministry of Education, Northeast Normal University, Changchun 130024, China}

\author{Tai-Yu Zheng\footnote{zhengty@nenu.edu.cn}}
\affiliation{Center for Quantum Sciences and School of Physics, Northeast Normal University, Changchun, 130024, People's Republic of China}
\affiliation{Center for Advanced Optoelectronic Functional Materials Research, and Key Laboratory for UV Light-Emitting Materials and Technology
of Ministry of Education, Northeast Normal University, Changchun 130024, China}

\date{\today}

\begin{abstract}
The Knill-Laflamme-Milburn (KLM) states have been proved to be a useful resource for quantum information processing [E. Knill, R. Laflamme, and G. J. Milburn, \href{http://dx.doi.org/10.1038/35051009}{Nature 409, 46 (2001)}]. For atomic KLM states, several schemes have been put forward based on the time-dependent unitary dynamics, but the dissipative generation of these states has not been reported. This work discusses the possibility for creating different forms of bipartite KLM states in neutral atom system, where the spontaneous emission of excited Rydberg states, combined with the Rydberg antiblockade mechanism, is actively exploited to engineer a steady KLM state from an arbitrary initial state. The numerical simulation of the master equation signifies that a fidelity above 99\% is available with the current experimental parameters.
\end{abstract}

\maketitle

\section{Introduction}
It is acknowledged that the generation and stabilization of quantum entanglement \cite{PhysRev.47.777,pra022317ref1} is a remarkable research field in quantum information science, which has various practical applications in quantum cryptography \cite{pra032323ref8}, quantum superdense coding \cite{pra032323ref9}, and quantum teleportation \cite{pra032323ref10}. As a specific class of entangled multiphoton states, the Knill-Laflamme-Milburn (KLM) can get over the issue that the success probability decreases with increasing complexity of the quantum computational scheme \cite{nature409}. { For example, utilizing the KLM state, the scalable quantum computation can be performed with the success probability $(1-1/n)$ where $n$ is the
number of ancilla. This value is asymptotically close to unity for a large number $n$, which is a sharp contrast to the success probability of $25\%$ due to the impossibility of performing complete Bell measurement \cite{prl110503ref3,prl110503}. Further more, based on the scheme of \cite{nature409}, Franson {it et al.} realized a high-fidelity quantum logic operations assisted by a more general KLM state, where each component of the quantum state is no longer equal weight superposition \cite{prl137901}. In their approach, the logic devices always produce an output with an intrinsic error rate proportional to $1/n^2$.}

Recently, researchers pay close attention to preparing and extending photonic KLM states \cite{cheng2012ref13,cheng2012ref15,cheng2012ref16,jpb195501}. Concretely, Franson \textit{et al.} made use of elementary linear-optics gates and solid-state approach to produce the arbitrary photon-number
KLM states \cite{cheng2012ref13}. In addition, Lemr {\it et al.} proposed two ways for
preparation of the KLM state using spontaneous
parametric down-conversion in experiment and employing a tunable controlled phase gate in theory, respectively \cite{cheng2012ref16,jpb195501}.
Nevertheless, all the schemes mentioned above are probabilistic due to the nature of the linear-optics system. { In \cite{cheng2012ref14},
Popescu proved that a KLM type quantum computation can be performed not only with photons but also with bosonic neutral atoms or bosonic ions, and this idea makes the atomic KLM state meaningful. On the other hand, by virtue of the cavity quantum electrodynamics technology
which can coherently couple atoms and photons, a deterministic KLM state for photons can
be obtained through the mapping between atoms and photons, which then can be used for photonic
quantum KLM computation.}

Up to present, the generation of KLM states for atoms or artificial atoms have been put forward with the time-dependent unitary dynamics \cite{cheng2012generation,liu2014preparation}. Although their works are deterministic, the decoherence effect arising from the weak interaction between quantum system and its surrounding environment would decrease the fidelity of the target quantum state.
Fortunately, there are also several schemes suggesting to prepare entangled states by dissipation \cite{prl090502ref12,prl090502ref13,prl090502ref11,pra012319ref13}, in which the dissipation can be tuned into a resource for certain quantum information process task. For instance,  Verstraete \textit{et al.} showed the opposite effect of dissipation on engineering a large variety of strongly correlated states in steady state \cite{prl090502ref11}. Kastoryano \textit{et al.} proposed a scheme for the preparation of a maximally entangled state of two atoms in an optical cavity. The cavity decay was no longer undesirable, but played an integral part in the dynamics \cite{pra012319ref14}. Very recently, Shao \textit{et al.} put forward a scheme for generating maximally entanglement between two Rydberg atoms through the atomic spontaneous emission \cite{PhysRevA.95.062339}.

{ The Rydberg atoms has been considered as a good candidate for quantum information processing due to the Rydberg blockade effect \cite{pra063419ref1,pra063419ref2,pra063419ref3}.}
The large size and the large electric dipole moment of Rydberg atoms can induce a long-range interaction. When the atoms are excited into Rydberg states in a small volume, the long-range interaction will prevent the multiple Rydberg excitations by the level shifts and can facilitate the formation of strongly correlated many-body systems.
The Rydberg blockade mechanism provides a promising method to create quantum states in applications ranging from quantum information
processing \cite{pra063419ref6,pra063419ref5} to quantum nonlinear optics \cite{pra063419ref8,pra063419ref9}.  However, Ates \textit{et al.} predicted an
opposite effect, namely, the Rydberg antiblockade \cite{pra012328ref18}: The two-photon detuning can compensate the energy shift of Rydberg states, which results in the simultaneous transitions of two atoms into Rydberg states and the  inhibition of Rydberg blockade. Since then, the Rydberg antiblockade has been proposed to deterministically implement quantum entanglement and multiqubit logic gates \cite{pra012328ref20,PhysRevA.89.012319,pra012328,pra022319,pra032336}. Particularly, Carr \textit{et al.} utilized Rydberg mediated interactions and dissipation to prepare the high fidelity entanglement and antiferromagnetic states \cite{pra012328ref20}. Su \textit{et al.} proposed an alternative scheme to quickly achieve the Rydberg antiblockade regime, which can be used to construct two- and multiqubit quantum logic gates \cite{pra022319}.

In this paper, we consider a dissipative scheme to prepare the bipartite KLM state via Rydberg antiblockade mechanism. We compensate the energy shift of Rydberg states by the two-photon detuning and the Stark shifts induced by classical optical lasers, making the atomic spontaneous emission as a powerful resource to create the KLM state, and eliminate the undesired states by a dispersive microwave field. Combined with these operations, the target state becomes the unique steady state of system. Therefore, the high fidelity entanglement can be realized without states initialization and the precisely evolution time.

\section{The full and effective Markovian master equations of system}
\begin{figure}[htbp]
\centering
\includegraphics[scale=0.16]{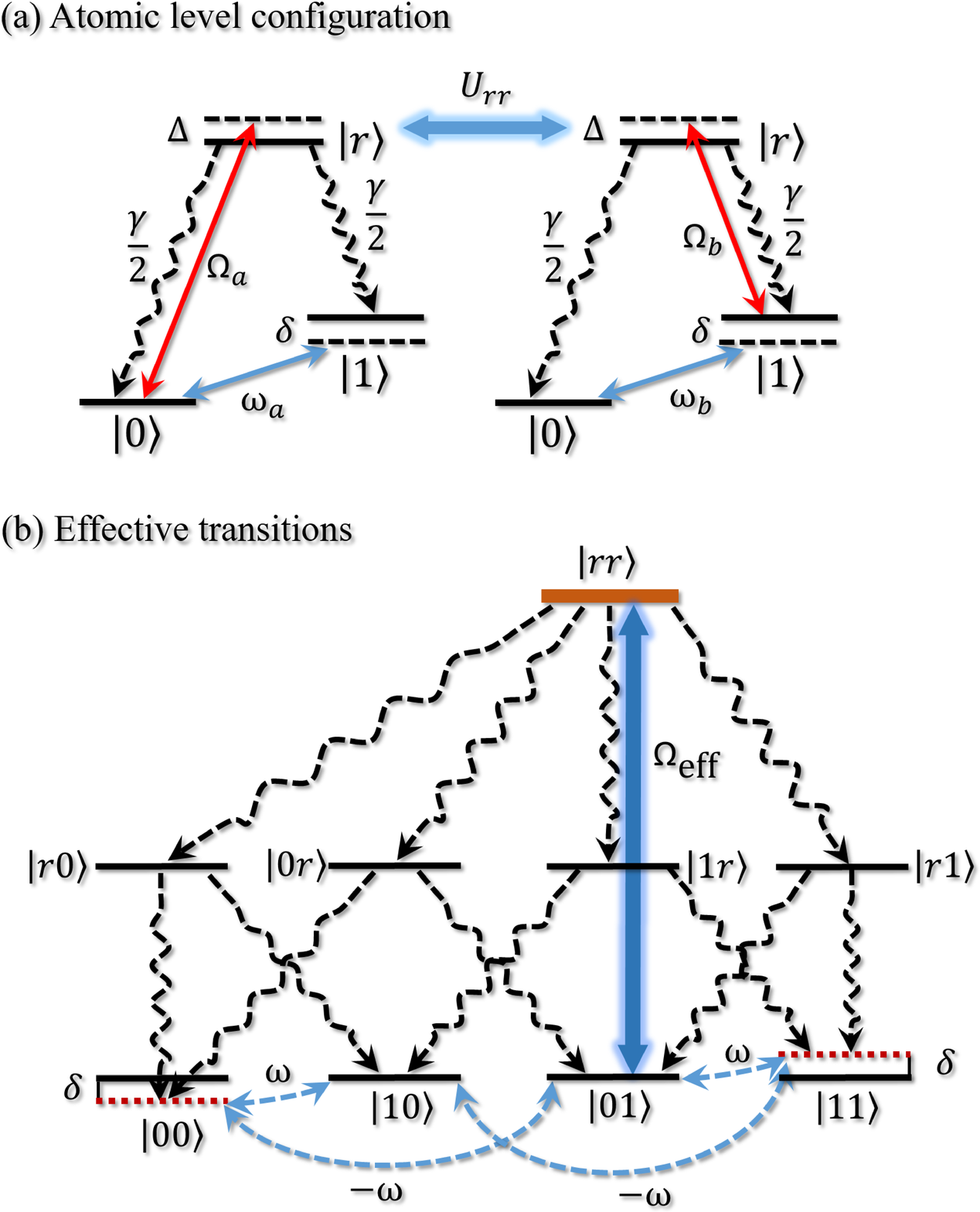}
\caption{(a) Diagrammatic illustration of the dissipative scheme and the KLM state is $|E_1\rangle=(|00\rangle+|10\rangle+|11\rangle)/\sqrt{3}$. (b) The effective transitions of the reduced system.}
\label{model}
\end{figure}
The KLM state  is defined as $|E_1\rangle=\sum_{j=0}^n|1\rangle^j|0\rangle^{n-j}/\sqrt{1+n}$ \cite{nature409}, where the notation $|a\rangle^j$ means $|a\rangle|a\rangle...,j$ times. For the case of $n = 2$, the bipartite KLM state reads
\begin{eqnarray}
|E_1\rangle=(|00\rangle+|10\rangle+|11\rangle)/\sqrt{3}.
\end{eqnarray}
To prepare the bipartite KLM state dissipatively, we show the atomic configuration in Fig.~\ref{model}(a), which illustrates two three-level Rydberg atoms consisting of two ground states $|0\rangle,|1\rangle$ and a Rydberg state $|r\rangle$. For the first atom, the ground state $|0\rangle$ is dispersively coupled to the excited state $|r\rangle$ by a classical field with Rabi frequencies $\Omega_a$, while the transition $|1\rangle\leftrightarrow|r\rangle$ of the second atom is dispersively driven by another classical field with Rabi frequencies $\Omega_b$. The driving fields are both detuned by $-\Delta$ and we set $\Omega_a=\Omega_b=\Omega$ for simplicity. The transitions between the ground states $|0\rangle$ and $|1\rangle$ are driven by two microwave fields of Rabi frequency $\omega_a=-\omega_b=\omega$, detuning $\delta$. The Rydberg state $|r\rangle$ is assumed to spontaneously decay down to the ground states $|0\rangle$ and $|1\rangle$ with an equal rate $\gamma/2$, respectively. The long-range interactions of
Rydberg atoms are described by $U_{rr}$ as both atoms occupy the same Rydberg states. In the interaction picture, the Hamiltonian of system can be written as
\begin{eqnarray}\label{HI}
H_I&=&\Omega e^{-i\Delta t}(|r\rangle_1\langle0|+|r\rangle_2\langle1|)+\sum_{j=1,2}\omega_j e^{i\delta t}|1\rangle_j\langle0|+{\rm H.c.}+U_{rr}|rr\rangle\langle rr|,
\end{eqnarray}
where the Rydberg-mediated interaction $U_{rr}$ originates from the dipole-dipole potential of the scale
$C_3/r^3$ or the long-range van der Waals interaction proportional to $C_6/r^6$, with $r$ being the distance
between two Rydberg atoms and $C_{3(6)}$ depending on the quantum numbers of the Rydberg
state \cite{PhysRevA.77.032723,RevModPhys.82.2313}.
After rotating the frame of the Hamiltonian, we obtain
\begin{eqnarray}\label{HIrot}
H_I&=&\Omega(|r\rangle_1\langle0|+|r\rangle_2\langle1|)+\omega(|1\rangle_1\langle0|-|1\rangle_2\langle0|)+{\rm H.c.}\nonumber\\
&&-\Delta(|r\rangle_1\langle r|+|r\rangle_2\langle r|)+\delta(|1\rangle_1\langle1|-|0\rangle_2\langle0|)+U_{rr}|rr\rangle\langle rr|.
\end{eqnarray}
The atomic spontaneous emission can be described by the Lindblad operators as $L^{1(2)}=\sqrt{\gamma/2}|0(1)\rangle_1\langle r|$, $L^{3(4)}=\sqrt{\gamma/2}|0(1)\rangle_2\langle r|$.  The evolution of the system is now governed by the Markovian master equation
\begin{eqnarray}\label{fullmaster}
\dot\rho=-i[H_I,\rho]+\sum_{i=1}^4L^i\rho L^{i\dag}-\frac{1}{2}(L^{i\dag}L^i\rho+\rho L^{i\dag}L^i).
\end{eqnarray}
According to the standard second-order perturbation theory, in the regime of the large detuning limit $\Delta\gg\Omega$ and $U_{rr}\sim2\Delta$, the process of the single excitation may be adiabatically eliminated and the two atoms can be directly excited into the Rydberg states simultaneously \cite{PhysRevA.95.062339}. Then the Hamiltonian can be rewritten as
\begin{eqnarray}
H_{\rm eff}&=&
\frac{2\Omega^2}{\Delta}|rr\rangle\langle01|+\omega(|00\rangle-|11\rangle)(\langle10|-\langle01|)+{\rm H.c.}\nonumber\\
&&+\delta(|11\rangle\langle11|-|00\rangle\langle00|)+\frac{\Omega^2}{\Delta}(|0\rangle_1\langle0|+|1\rangle_2\langle1|)\nonumber\\
&&+(U_{rr}-2\Delta+\frac{2\Omega^2}{\Delta})|rr\rangle\langle rr|.
\end{eqnarray}
The Stark shifts $\Omega^2/\Delta(|0\rangle_1\langle0|+|1\rangle_2\langle1|)$ can be canceled by introducing other ancillary levels and the term $2\Omega^2/\Delta|rr\rangle\langle rr|$ can be counteracted by setting $U_{rr}=2\Delta-2\Omega^2/\Delta$.  Then we reformulate the above Hamiltonian as follows
\begin{eqnarray}\label{Heff}
H_{\rm eff}&=&\Omega_{\rm eff}|rr\rangle\langle01|+\omega(|00\rangle-|11\rangle)(\langle10|-\langle01|)+{\rm H.c.}\nonumber\\
&&+\delta(|11\rangle\langle11|-|00\rangle\langle00|),
\end{eqnarray}
where $\Omega_{\rm eff}=2\Omega^2/\Delta$.

{ The corresponding effective Lindblad operators are
\begin{eqnarray}\label{Leff}
L_{\rm eff}^{1(2)}&=&\sqrt{\frac{\gamma}{2}}|r0(1)\rangle\langle rr|,L_{\rm eff}^{3(4)}=\sqrt{\frac{\gamma}{2}}|0(1)r\rangle\langle rr|,\\
L_{\rm eff}^{5(6)}&=&\sqrt{\frac{\gamma}{2}}|0(1)0\rangle\langle r0|,L_{\rm eff}^{7(8)}=\sqrt{\frac{\gamma}{2}}|0(1)1\rangle\langle r1|,\\
L_{\rm eff}^{9(10)}&=&\sqrt{\frac{\gamma}{2}}|00(1)\rangle\langle 0r|,L_{\rm eff}^{11(12)}=\sqrt{\frac{\gamma}{2}}|10(1)\rangle\langle 1r|.
\end{eqnarray}
Finally we have the effective master equation as
\begin{eqnarray}\label{effmaster}
\dot\rho=-i[H_{\rm eff},\rho]+\sum_{k=1}^{12}L_{\rm eff}^k\rho L_{\rm eff}^{k\dag}-\frac{1}{2}(L_{\rm eff}^{k\dag}L_{\rm eff}^k\rho+\rho L_{\rm eff}^{k\dag}L_{\rm eff}^k).
\end{eqnarray}
According to Eq.~(\ref{effmaster}), we plot the effective transitions of the reduced system in Fig.~\ref{model}(b) to explain how the scheme works. The system consists of four ground states $|00\rangle, |10\rangle, |01\rangle, |11\rangle$, four mediate states $|r0\rangle,|r1\rangle,|0r\rangle,|1r\rangle$ and an excited state $|rr\rangle$. The four ground states can be driven to each other by the microwave fields. (The effect of the microwave fields nearly contributes nothing to the dissipative dynamics of the mediate state since they are unstable, which can be neglected in the effective master equation.)
And the ground state $|01\rangle$ can also be coupled to the excited state $|rr\rangle$ with an effective Rabi frequency $\Omega_{\rm eff}$. Due to the spontaneous emission of atoms, the excited state $|rr\rangle$ will decay to the four mediate states and further decay to the four ground states by atomic spontaneous emission. Thus the
 system will repeat the process of pumping and decaying. We find that when the detuning of microwave field satisfies $\delta=\omega$, the bipartite KLM state turns into the unique steady-state solution to the Eq.~(\ref{effmaster}) because of $H_{\rm eff}|E_1\rangle=0$ and $L_{\rm eff}^k|E_1\rangle=0$. Consequently, the system finally  will be stabilized at the bipartite KLM state.}

\section{The demonstration of the validity of effective master equation and the investigation of relevant parameters}
\begin{figure}[h]
\centering
\includegraphics[scale=0.5]{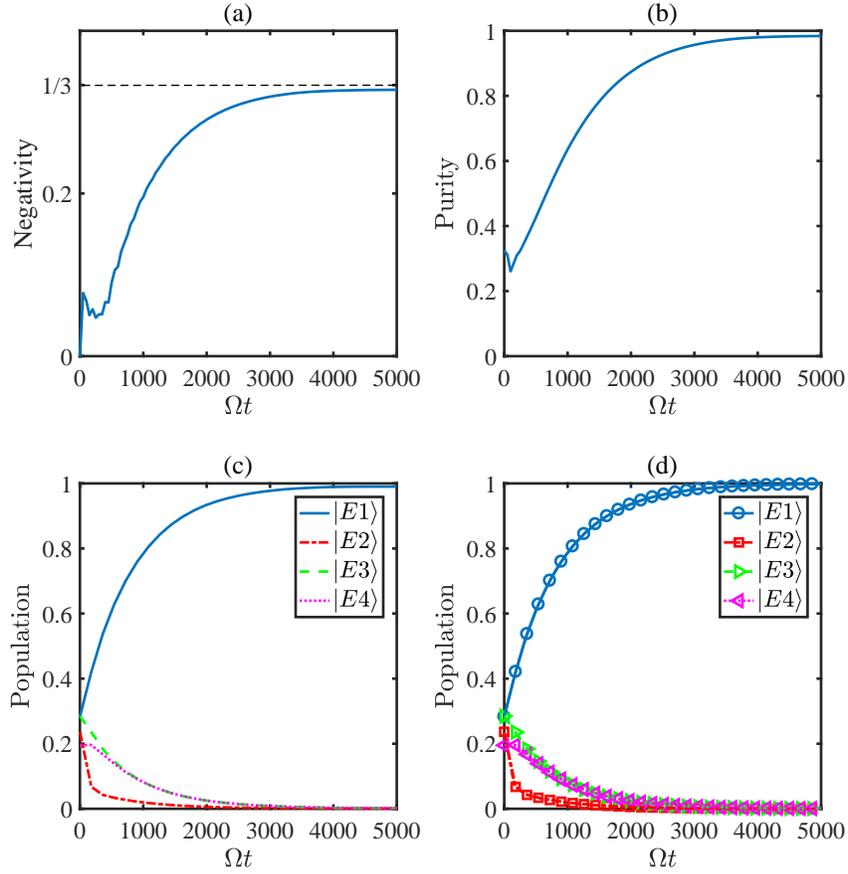}
\caption{\label{zong} {(a) The evolutions of the negativity of system governed by the full master equation. (b) The evolution of the purity of system, Tr$[\rho^2(t)]$, governed by the full master equation. (c) and (d) show the populations of states $|E_1\rangle,|E_2\rangle,|E_3\rangle$ and $|E_4\rangle$ as functions of $\Omega t$ governed by the full and effective master equations, respectively. For all figures, the initial states are chosen arbitrarily as $\rho_0=a|00\rangle\langle00|+b|11\rangle\langle11|+c|10\rangle\langle10|+d|01\rangle\langle01|$, where $a=0.3,~b=0.15,~c=0.45,$ and $d=0.1$.  The relevant parameters: $\Delta=70\Omega,~\delta=0.02\Omega,$ and $\gamma=0.05\Omega$.}}
\end{figure}

{ In order to fully demonstrate the validity of our proposal, we measure the evolution of the system with different quantities. In Fig.~\ref{zong}(a), we investigate the negativity of state $\rho(t)$ solved by the full master equation as a function of $\Omega t$. The negativity is a measure proposed by Vidal \textit{et al.} \cite{pra032314}. It can bound two relevant quantities, the channel capacity and
the distillable entanglement, characterizing the entanglement of mixed states. The definition of negativity is
\begin{equation}
\mathcal{N}(\rho_{A,B})=\frac{||\rho_{A,B}^{T_A}||-1}{2},
\end{equation}
where
\begin{equation}
||\rho_{A,B}^{T_A}||={\rm Tr}\sqrt{\rho_{A,B}^{T_A\dag}\rho_{A,B}^{T_A}},
\end{equation}
and $\rho_{A,B}^{T_A}$ denotes the partial transpose of the bipartite mixed state $\rho$ on the subsystem $A$. To increase the credibility, the initial state is chosen arbitrarily as $\rho_0=a|00\rangle\langle00|+b|11\rangle\langle11|+c|10\rangle\langle10|+d|01\rangle\langle01|$, where $a=0.3,~b=0.15,~c=0.45,$ and $d=0.1$. We can obtain the negativity of $\rho(t)$ governed by full master equation can reach $0.3275$ at $\Omega t=5000$, which approaches the ideal value of the negativity of the bipartite KLM state $1/3$. However, we cannot determine the steady state is the pure KLM state at this stage because other mixed states may have the same negativity.

In Fig.~\ref{zong}(b), we further estimate the evolution of system by purity, which is defined as $\mathcal{P}(t)={\rm Tr}[\rho^2(t)]$. From the definition, we can know that once the system is in a pure state, the purity will be equal to unit, otherwise, the purity will be less than unit. In Fig.~\ref{zong}(b), the system also begins with the same mixed state $\rho_0=a|00\rangle\langle00|+b|11\rangle\langle11|+c|10\rangle\langle10|+d|01\rangle\langle01|$, with $a=0.3,~b=0.15,~c=0.45,$ and $d=0.1$. At $\Omega t=5000$, the purity can achieves $0.985$. Combined with the negativity in Fig.~\ref{zong}(a), the behavior of Fig.~\ref{zong}(b) can indirectly show that the bipartite KLM state is produced and directly show that the system tends to be a pure state.

\begin{figure*}
\begin{minipage}[t]{0.49\linewidth}
\centering
\includegraphics[scale=0.44]{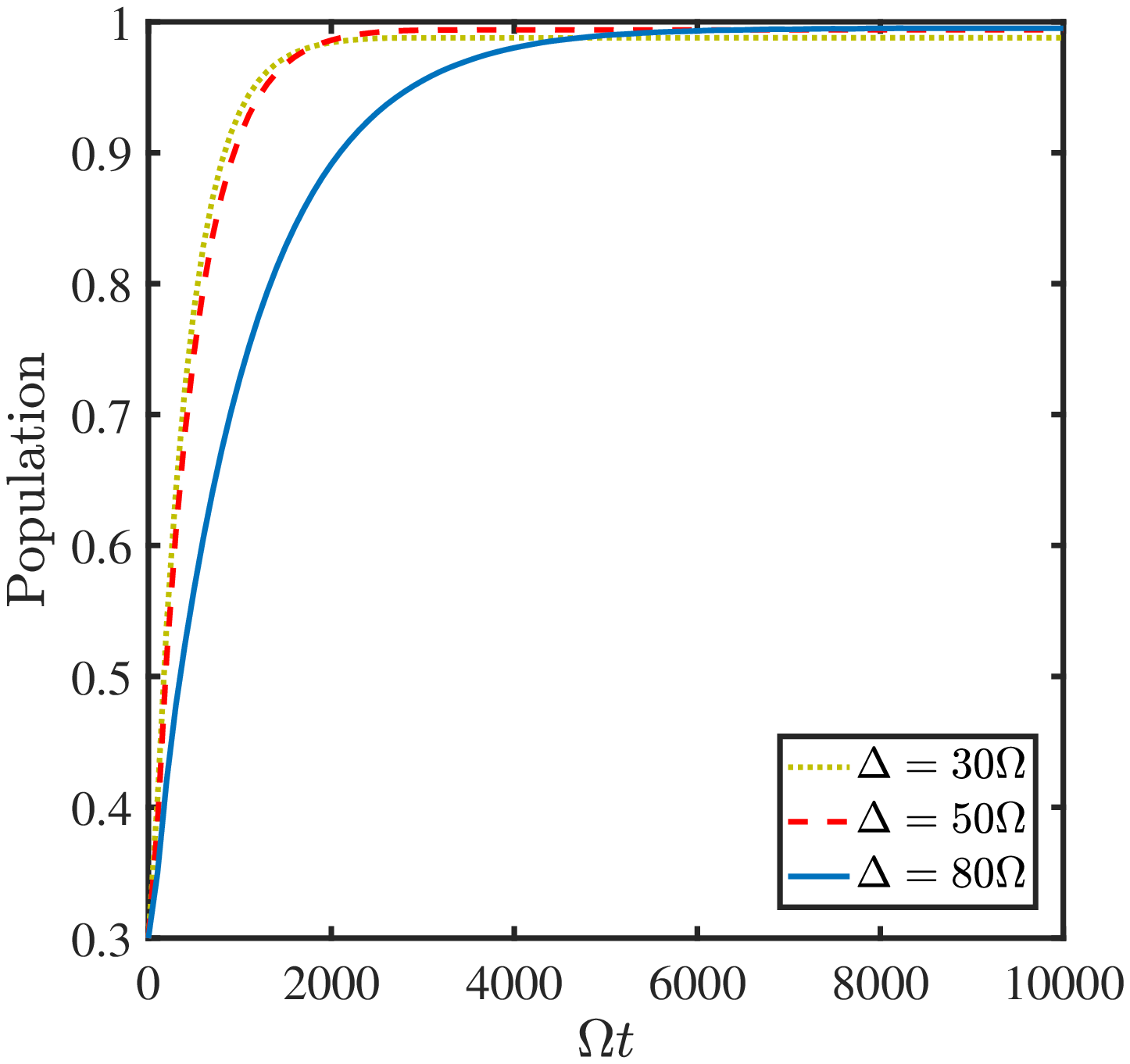}
\centerline{(a)}
\end{minipage}
\begin{minipage}[t]{0.49\linewidth}
\centering
\includegraphics[scale=0.44]{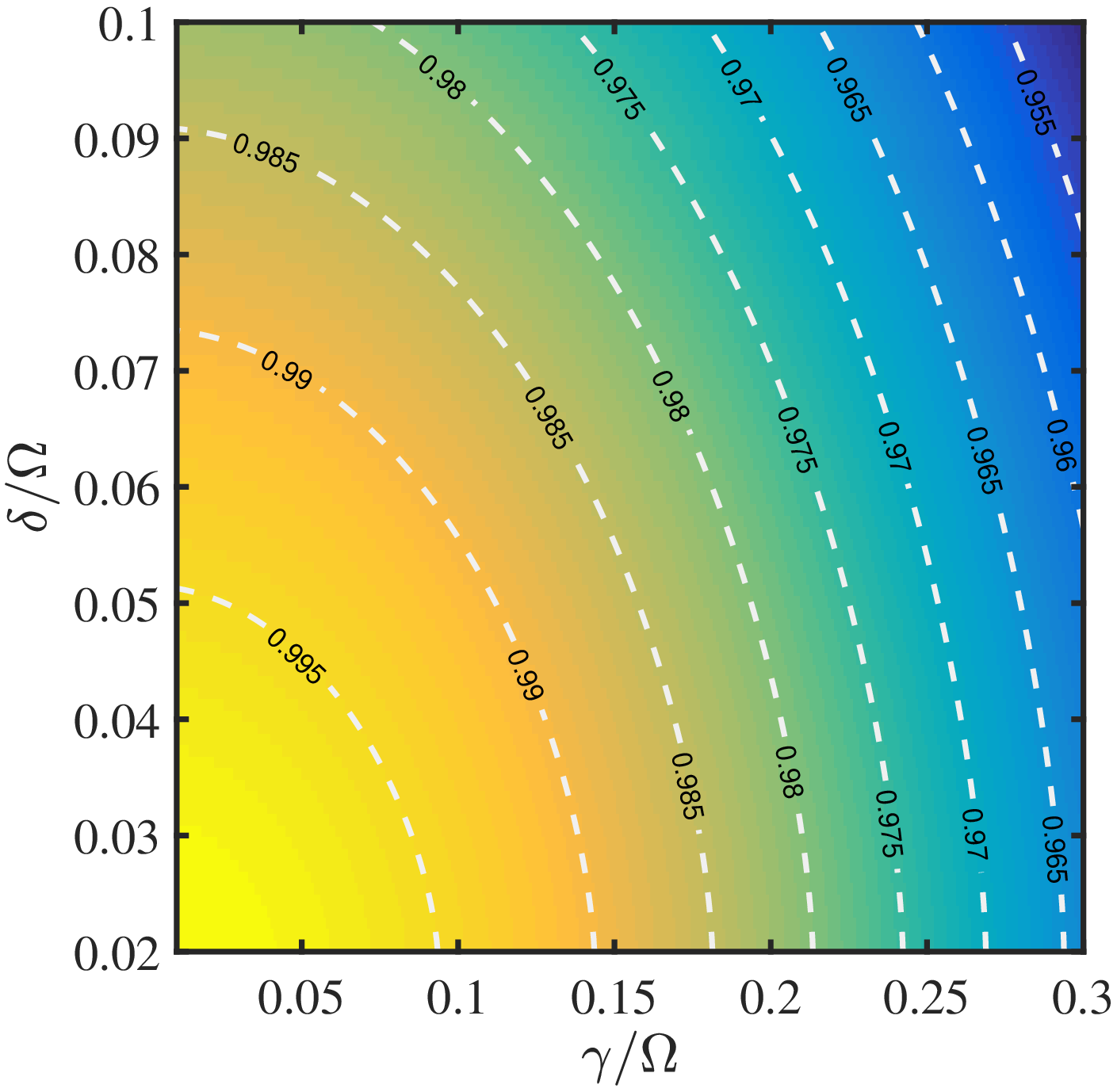}
\centerline{(b)}
\end{minipage}
\caption{\label{Delta} (a) The population of the KLM state $|E_1\rangle$ as a function of $\Omega t$ with different detuning parameters $\Delta$. The initial state is the same as that of Fig.~\ref{zong}. The relevant parameters are chosen as $\delta=0.02\Omega,$ and $\gamma=0.05\Omega$. (b) Contour plot (dashed lines) of the steady-state fidelity of the bipartite KLM state, where the fidelity is defined as $F={\rm Tr}[\sqrt{\sqrt{\rho_E}\rho(t)\sqrt{\rho_E}}]$ and we have set $\Delta=50\Omega$.}
\end{figure*}

In the Fig.~\ref{zong}(c) and Fig.~\ref{zong}(d), we plot the evolutions of the population for state $|\varphi\rangle$ governed by the full and effective master equations, respectively. The population of state $|\varphi\rangle$ is defined as $P=\langle \varphi|\rho(t)|\varphi\rangle$. In the Fig.~\ref{zong}(c), it convincingly demonstrates the feasibility of the present scheme that the population of target state $|E_1\rangle$ (solid line) tends to unity and the other orthonormal states $|E_2\rangle$ (dotted-dashed line), $|E_3\rangle$ (dashed line), and $|E_4\rangle$ (dotted line) tend to vanish, where
\begin{eqnarray}
|E_2\rangle&=&\frac{1}{\sqrt{15}}(|00\rangle-3|01\rangle-2|10\rangle+|11\rangle),\\
|E_3\rangle&=&\frac{1}{\sqrt{5}}(\theta_+|00\rangle+|01\rangle-|10\rangle-\theta_-|11\rangle),\\
|E_4\rangle&=&\frac{1}{\sqrt{5}}(\theta_-|00\rangle-|01\rangle+|10\rangle-\theta_+|11\rangle),
\end{eqnarray}
and $\theta_\pm=(\sqrt{5}\pm1)/2$. In Fig.~\ref{zong}(d), the populations of the reduced system are in good agreement with the corresponding populations in Fig.~\ref{zong}(c), which proves the validity of the effective system and the above physical explanation of Fig.~\ref{model}(b).}

Now we investigate the influence of the detuning parameter $\Delta$ on the generation of the KLM state in Fig.~\ref{Delta}(a). According to the antiblockade of Rydberg atoms, we need the large detuning limit $\Delta\gg\Omega$ to adiabatically eliminate the process of the single excitation. In Fig.~\ref{Delta}(a), the limiting condition is destroyed by the decrease of $\Delta$, which leads to the reduction of the population for the target state. But the convergence time will be shorter with the decrease of $\Delta$, since the effective coupling  strength $\Omega_{\rm eff}=2\Omega^2/\Delta$ is enlarged.

Another method to evaluate the quality of the steady sate is the calculation of the fidelity. In Fig.~\ref{Delta}(b), we give the steady-solution of system by solving the full master equation $\dot\rho=0$, and study the steady-state fidelity of the bipartite KLM state as functions of atomic spontaneous emission and the detuning of the microwave field, where the steady-state fidelity is defined as $F={\rm Tr}[\sqrt{\sqrt{\rho_E}\rho(t)\sqrt{\rho_E}}]$. The dashed lines represent some contours of the fidelity. { The figure makes an impactful proof that the present scheme can realize a high-fidelity bipartite KLM state with a wide range of relevant parameters: When $\gamma/\Omega=0.3$ and $\delta/\Omega=0.1$, the fidelity of target state is still above $95\%$.}

\section{Experimental feasibility}
\begin{figure}
\centering
\includegraphics[scale=0.16]{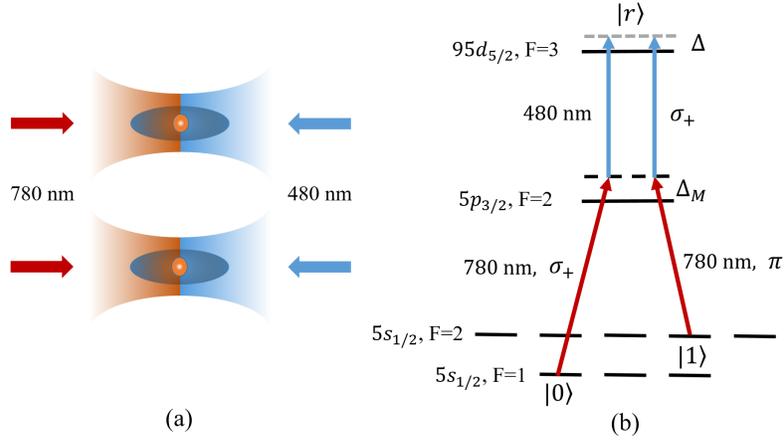}
\caption{\label{exsetup}(a) Scheme of the experimental setup. Two ${}^{87}$Rb atoms are driven by two laser beams, respectively.   (b) Internal energy levels of the corresponding atoms.}
\end{figure}

{  In experiment, we may employ $^{87}$Rb atoms in our proposal as shown in Fig.~\ref{exsetup}. Each atom can be addressed individually by two classical fields.
The transition from the ground state $|0\rangle=|5s_{1/2};F=1\rangle$ to the mediated state $|5p_{3/2};F=2\rangle$ of the first atom is driven by a $\sigma_+$-polarized laser beam at $780$ nm, detuned by $-\Delta_M$. A $\sigma_+$-polarized $480$-nm laser beam achieves the further transition from the mediated state $|5p_{3/2};F=2\rangle$ to the Rydberg state $|r\rangle=|95d_{5/2};F=3\rangle$ with detuning $\Delta_M-\Delta$. For the second atom, a $\pi$-polarized $780$-nm laser beam couples the ground state $|1\rangle=|5s_{1/2};F=2\rangle$ to the mediated state $|5p_{3/2};F=2\rangle$ with detuning $-\Delta_M$, and a $\sigma_+$-polarized $480$-nm laser beam then couples the excited state  $|5p_{3/2};F=2\rangle$ to the the Rydberg state $|r\rangle=|95d_{5/2};F=3\rangle$ with detuning $\Delta_M-\Delta$. The detuning paremeter $\Delta_M$ is made large enough compared with the Rabi frequencies of driving fields that the population of the mediated state can be safely omitted \cite{PhysRevLett.104.010502,PhysRevLett.104.010503,PhysRevA.82.030306}. The coupling between two ground states can be directly completed by microwave fields or equivalent Raman transition, which is not shown in Fig.~\ref{exsetup}.}

\begin{figure}
\centering
\includegraphics[scale=0.5]{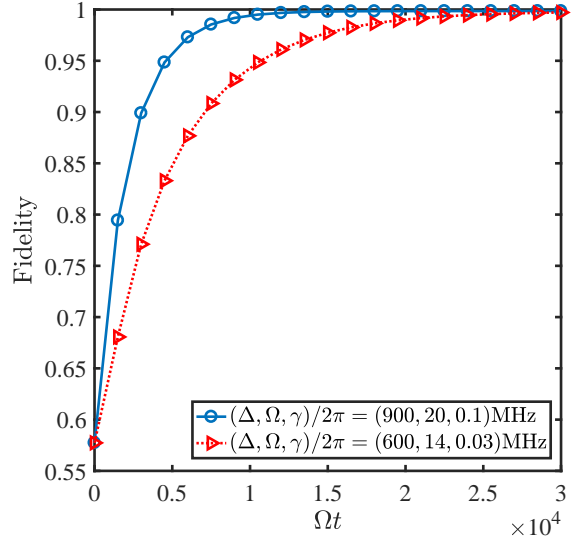}
\caption{\label{extime}The fidelity of the KLM state $|E_1\rangle$ as a function of $\Omega t$ with different experimental parameters and $\delta=0.02\Omega$. The initial states are all $\rho_0=|00\rangle\langle00|$. }
\end{figure}

The
Rabi laser frequency $\Omega$ realized by the above two-photon process can be tuned continuously between $2\pi\times(0,60)~$MHz \cite{prl090402}. Thus, considering the decay rate of the Rydberg state $\gamma=2\pi\times0.03~$MHz \cite{njp043020}, we choose the parameters $\Omega=14$~MHz and $\Delta=600$~MHz, and obtain the fidelity of $99.67\%$ { (empty triangle)} at $t=30000/\Omega\approx2.14$~ms in Fig.~\ref{extime}. For another configuration of Rydberg atoms, such as in \cite{prl223002} where the principal quantum number of Rydberg state of  $^{87}$Rb atom are chosen as $N=20$, the decay rate can reach $2\pi\times100~$kHz. In this situation, a choice of $\Omega=20$~MHz and $\Delta=900$~MHz will guarantee a steady-state fidelity $99.70\%$ at $t=1.5$~ms, as indicated by the {  empty circle} in Fig.~\ref{extime}. To sum up, we are always able to achieve a high fidelity by selecting different Rabi frequencies and detuning parameters with regard to different decay rates for the excited Rydberg states.
{\section{Generalization: Preparation of a more general bipartite KLM state }
{ The general KLM state (triangle-shaped amplitude function), $|E_1'\rangle=\alpha_0|00\rangle+\alpha_1|10\rangle+\alpha_0|11\rangle,$ can be used to realize a high-fidelity approach to linear optics quantum computing, which does not rely on postselection, and one can choose the suitable values of $\alpha_j$ to significantly increase the efficiency of teleportation based quantum computing \cite{prl137901,jpb195501}. In our scheme, we can change the relation between the Rabi frequency $\omega$ and detuning $\delta$ of microwave fields to create a general bipartite KLM state.
For instance, we now set $\delta=m\omega$ and calculate the steady solution of the master equation by $\dot\rho=0$. The unique steady-state solution of the system reads
 \begin{eqnarray}
|E_1'\rangle=\frac{1}{\sqrt{2+m^2}}|00\rangle+\frac{m}{\sqrt{2+m^2}}|10\rangle+\frac{1}{\sqrt{2+m^2}}|11\rangle,
\end{eqnarray}
which is just a general bipartite KLM state. The other orthonormal ground states can be written as
\begin{eqnarray}
|E_2'\rangle&=&\frac{1}{\sqrt{(2+m^2)(4+m^2)}}\Big[m|00\rangle-(2+m^2)|01\rangle-2|10\rangle+m|11\rangle\Big],\\
|E_3'\rangle&=&\frac{1}{\sqrt{4+m^2}}(\theta_+'|00\rangle+|01\rangle-|10\rangle-\theta_-'|11\rangle),\\
|E_4'\rangle&=&\frac{1}{\sqrt{4+m^2}}(\theta_-'|00\rangle-|01\rangle+|10\rangle-\theta_+'|11\rangle),
\end{eqnarray}
where $\theta_\pm'=(\sqrt{3+2m^2}\pm1)/2$.
\begin{figure}
\centering
\includegraphics[scale=0.5]{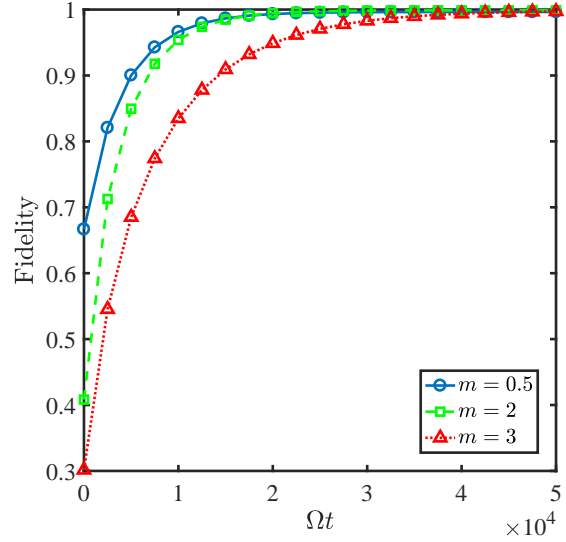}
\caption{\label{mchange}The fidelity of the general bipartite KLM state $|t_2\rangle$ as a function of $\Omega t$ with different $m$. The relevant parameters are chosen as the experimental parameters $(\Delta,\Omega,\gamma)/2\pi=(900,20,0.1)$MHz and $\delta=0.02\Omega$. The initial states are all in $\rho_0=|00\rangle\langle00|$. }
\end{figure}

In order to demonstrate the success of the generalization, we plot the fidelity of the general bipartite KLM state $|E_1'\rangle$ as a function of $\Omega t$ with different $m$ in Fig.~\ref{mchange}. Without loss of generality, we select one set of experimental parameters of Fig.~\ref{extime}, i.e. $(\Delta,\Omega,\gamma)/2\pi=(900,20,0.1)$MHz and $\delta=0.02\Omega$. We can find that when $\Omega t=5\times10^4$, the lines of $m=0.5$ (empty circle), $m=2$ (empty square) and $m=3$ (empty triangle) can reach $99.64\%$, $99.88\%$ and $99.72\%$, respectively. These results fully prove the validity of the generalization.}

\section{Summary}
In summary, we have systematically investigated the feasibility of dissipatively generating the bipartite KLM state by two three-level Rydberg atoms. The atomic spontaneous emission acts as a powerful resource to create the bipartite KLM state.  We counteract the energy shift of Rydberg states by the two-photon detuning, adiabatically eliminate the process of the single excitation by the large detuning limit, and pick up the desired state by dispersive microwave fields. Thus, the target state is the unique steady state of system, and it can be generated in a wide range of relevant parameters. Finally, we generalize our scheme to prepare a more general bipartite KLM state by adjusting the relation between the Rabi frequency and detuning of microwave fields. The numerical simulation reveals that both the standard KLM state and the generalized KLM state process high fidelities with present experimental parameters.

\section*{Funding}
National Natural Science Foundation of China (NSFC) (11534002, 61475033, 11774047); Fundamental Research Funds for the Central Universities (2412016KJ004).

\bibliographystyle{apsrev4-1}
\bibliography{klm}

\end{document}